\documentclass[12pt,oneside,english]{article}
\usepackage[T1]{fontenc}
\usepackage[latin1]{inputenc}
\usepackage{babel}
\usepackage{setspace}
\makeatletter
\begin{document}

\section*{\bf  Lax-Phillips evolution as an evolution of 
Gell-Mann-Hartle-Griffiths 
 histories  and emergence  of the Schr\"oedinger
equation for a stable history} 
\makeatother

\bigskip \bigskip 

 \centerline{\large D.Bar$^{a}$ and L.P.Horwitz$^{a,b}$}  
  \bigskip 
  
{\bf $^a$Department of Physics, Bar Ilan University, Ramat Gan,
Israel.  \\ $^b$Raymond and Beverly Sackler Faculty of Exact Science, School of
Physics, Tel Aviv University, Ramat Aviv, Israel. }
\bigskip \bigskip 

\begin{abstract}
  { \it Using the  
 Gell-Mann-Hartle-Griffiths  formalism in the framework of  the Flesia-Piron 
 form of the 
 Lax-Phillips theory we show that the
Schr\"oedinger  equation may be derived as a condition of stability of 
histories. 
 This mechanism  is
realized in a mathematical structure closely related to the 
 Zeno effect.}
\end{abstract}
\bigskip  \bigskip 
\noindent
{PACS number(s): 03.65.-w, 03.65.Xp, 03.65.Db, 03.65.Bz. \\ \bigskip 
\bigskip \bigskip \noindent {\bf Keywords: Schr\"oedinger equation, Zeno effect,
Functional analysis.}} \par

\bigskip 

\bigskip \noindent 
 The possibility of  obtaining  physical effects  due to a large number of
 repetitions of the same measurement and interaction has been discussed both at
 the theoretical \cite{Zeno,Aharonov,Facchi} and experimental \cite{Itano}
 levels. Moreover, it has been argued \cite{Simonius} that these repetitions not only produce
 quantum effects, which is the reason for  calling them  quantum Zeno effect
 \cite{Zeno,Aharonov}, but they  may also be a source of macroscopic and 
 classical
 effects \cite{Bar}. \par We show here, using the functional Lax-Phillips (LP) 
 \cite{Lax} 
 generalization of
 quantum mechanics \cite{Horwitz,Strauss} and the histories formalism of
 Gell-Mann,  
 Hartle \cite{Hartle,Isham} and Griffiths \cite{Griffiths}  (GMHG) that for a special 
 choice of
evolution of histories one obtains stability of the GMHG state by a
mechanism that appears to be the limit of a large number of repetitive
measurements in a finite total time.  The stability of the GMHG state is
associated with the Schr\"odinger relation on the structure of the
histories, and hence characterizes the consistent subset. \par 
 The   
theory of Lax-Phillips \cite{Lax} which was originally formulated to describe resonances and semigroup evolutions
 (i.e., irreversible processes) for  the classical scattering of electromagnetic 
 waves on a finite target has been 
generalized to the quantum level by Flesia and Piron \cite{Horwitz}, 
 and Horwitz 
and Piron \cite{Horwitz} (see also \cite{Strauss}).  The appropriate space 
is  the "direct integral  Hilbert space" \cite{Horwitz,Gelfand} \begin{equation}\tilde H=\int_{t}\oplus 
H_{t}d\mu (t),  \label{e1}\end{equation} where we take the  measure $d\mu(t)$ 
 to be of  
 Stieljes-Lebesgue type, and $t$ corresponds to  a foliation parameter
playing the role of the time observable.  An 
element of $\tilde H$ is the sequence \cite{Strauss,Gelfand} (which we represent here as
countable) \begin{equation}\phi=
(h_{t_1},h_{t_2},h_{t_3}.....),   \label{e2}\end{equation} 
so that $\phi_{t_i}=h_{t_i}$ and  $h_{t_i} \subset {\cal H}_i$ where  the sequence 
${\cal H}_i$ corresponds 
 to a set of isomorphic {\it auxiliary}   Hilbert 
spaces \cite{Strauss,Gelfand} at the times $t_i$. Thus, we see that $\phi$ 
corresponds to a 
virtual history \cite{Strauss}. The evolution 
operator  $U(\tau )$ on $\tilde H$ 
is defined (for convenience here, on the continuum)  
\begin{equation}(U(\tau)\phi)_{t+\tau}=\phi^\tau_{t+\tau} =V_t(\tau)\phi_t, 
\label{e3} \end{equation}  where $V_t(\tau)$ is an  operator which is 
unitary on ${\cal H}_t$. The 
superscript $\tau$ on $\phi$ is the laboratory time which is a parameter, while $t$ is a 
dynamical variable. The subscript $(t+\tau)$ signifies that the original element 
from Eq (\ref{e2}) has been translated along the $t$ axis by $\tau$. That is, the action 
of the operator $U(\tau)$ has produced a  (auxiliary)   
Hilbert space unitary evolution combined with translation along the $t$ axis by 
the amount $\tau$. Note that the operator $U(\tau)$ forms a one-parameter group
\cite{Horwitz}, that is, 
$$(U(\tau_2)U(\tau_1)\phi)_{t+\tau_1+\tau_2}=
(U(\tau_1+\tau_2)\phi)_{t+\tau_1+\tau_2}$$  According to Eq (\ref{e3}) 
\begin{equation} \label{e4}  U(\tau)\phi =
(V_{t_1}(\tau)h_{t_1}, V_{t_2}(\tau)h_{t_2}\ldots)=
(h'_{t_1+\tau}, h'_{t_2+\tau} \ldots),
 \end{equation}
 where 
the primes indicate that the evolution constitutes a translation along the $t$
axis as well as a unitary evolution on the auxiliary Hilbert spaces.  We may think that 
 the   elements on the 
right hand side of the preceding equation have undergone a specific dynamical evolution 
which is only one of a great number of possible alternatives which may be
distinguished by appropriate projections. In that case we 
can represent each such element by  projection operators in the auxiliary space 
 at  specific times.  
Let us assume  that the initial state
$\phi_{\alpha}$ corresponds to a projected chain (denoted as $C_{\alpha}\phi$), i.e., 
\begin{equation} \label{e5} \phi_{\alpha}=(P_{\alpha_1}(t_1)h_{t_1},
P_{\alpha_2}(t_2)h_{t_2} \ldots ) \end{equation} Then  
\begin{eqnarray} &&  U(\tau)\phi_{\alpha}=(V_{t_1}(\tau)P_{\alpha_1}(t_1)h_{t_1},
V_{t_2}(\tau)P_{\alpha_2}(t_2)h_{t_2} \ldots )= \label{e6} \\ &&=(P_{\alpha_1}(t_1+\tau)
 h_{t_1+\tau},
P_{\alpha_2}(t_2+\tau) h_{t_2+\tau} \ldots ), \nonumber \end{eqnarray}
where \begin{equation} \label{e7}
P_{\alpha_i}(t_i+\tau)=V_{t_i}(\tau)P_{\alpha_i}(t_i)V^{-1}_{t_i}(\tau)
\end{equation} 
The unitary evolution of the projected history remains a projected history,
i.e., \begin{equation} \label{e8} U(\tau)C_{\alpha}\phi =C_{\alpha}(\tau)U(\tau)\phi,  
\end{equation}
where one may write \begin{equation} \label{e9} C_{\alpha}(\tau)=
U(\tau)C_{\alpha}U^{-1}(\tau) \end{equation} Thus, the Flesia-Piron generalized
state of Eq (\ref{e2}) may be considered as a GMHG history  developing
dynamically under a {\it unitary evolution of histories}. \par  
   In the GMHG  histories formalism \cite{Hartle,Isham,Griffiths}   
one  deals with a  set of alternative histories \cite{Hartle}, which 
are defined in the most simple 
example by giving sequences of projections at definite moments of time
$t_1,t_2,t_3.....t_n$.  The sequences  are denoted \cite{Hartle} 
by ${P^1_{\alpha_1}(t_1)},
{P^2_{\alpha_2}(t_2)},{P^3_{\alpha_3}(t_3)}.....{P^n_{\alpha_n}(t_n)}$. The 
projections may be   different at different times, for example, in the two slit 
experiment \cite{Feynman} ${P^2_{\alpha_2}(t_2)}$ could distinguish 
whether 
the electron went through the upper slit or the lower one at time $t_2$, while 
${P^3_{\alpha_3}(t_3)}$ might distinguish various places of arrival at the 
final screen at time $t_3$. In general, $\alpha_i$ corresponds to eigenstates of
a set of observables at $t_i$. Each set of $P's$ satisfies \cite{Hartle}  
\begin{equation} \sum_{\alpha_K}P^K_
{\alpha_K}(t_K)=I, \ \ \
P^K_{\alpha_K}(t_K)P^K_{\alpha'_K}(t_K)=\delta_{\alpha_K\alpha'_K}P^K_{\alpha_K}(t_K),  
\label{e10} \end{equation} 
indicating 
 that the $\alpha$'s  
 represent an exhaustive set of exclusive alternatives. An individual 
history corresponds to a particular sequence: $\alpha =(\alpha_1......\alpha_n)$ 
and for each history there is a corresponding chain of time ordered projection 
operators \cite{Hartle} 
$C_{\alpha}=P^n_{\alpha_n}(t_n).....P^1_{\alpha_1}(t_1)$. Such 
histories are termed coarse grained \cite{Hartle} when the $P's$ are not projections onto a 
basis (a complete set of states), and when there is not a set of $P's$ at each 
and every time,  otherwise, they are fine grained.   
When the initial state is 
pure 
one  can resolve it, by using the previous equations, into branches corresponding to 
the individual members of any set of alternative histories. That is, denoting in 
the Heisenberg picture  the initial state by $|\Psi\!>$ one
obtains \cite{Hartle}  \begin{equation}
|\Psi\!>=\sum_
{\alpha}C_{\alpha}|\Psi\!>=\sum_{\alpha_1....\alpha_n}P^n_{\alpha_n}(t_n).....
P^1_{\alpha_1}(t_1)|\Psi\!>  \label{e11} \end{equation} The vector $C_{\alpha}|\Psi\!>$ is the 
branch of $|\Psi\!>$ that corresponds to the individual history $\alpha$. \par 
If $\phi$, $\xi$ correspond to the GMHG histories ${\phi_k}$, ${\xi_k}$
respectively, the scalar product between them is given by \cite{Hartle,Isham}
(we shall use this definition below)
\begin{equation} \label{e12}
(\!\phi,\xi\!)=\prod_k(\!\phi_k,\xi_k\!)_{{\cal H}_{t_k}} \end{equation} 
If the set $\alpha$, $\grave \alpha$ differ by one projection $P_{\alpha_k}$ in
the sequence for which $P_{\alpha_k}^k(t_k)P_{\grave \alpha_k}^k(t_k) \\ =0$, then 
$<\!\Psi_{\alpha}|\Psi_{\grave \alpha}\!>=0$ where $|\Psi_{\grave \alpha}\!>$ is
the branch of $|\Psi\!>$ that corresponds to the history $\alpha$ (see Eq. 
\ref{e11})).  Following the definition given by Isham \cite{Isham} , embedding the space of
history filters in the orthocomplemented lattice of history propositions,
we may define a density matrix $\rho$ in term of {\it a priori}
probabilities  over the arbitrary histories  that form, in this space, a
complete set.   The density operator associated
 with  such  a  state is \begin{equation} \label{e13} 
\rho=\sum_{\alpha}p_{\alpha}\prod_{k=1}^{k=n}P_{\alpha_k}^k(t_k), \end{equation}
where the sum is taken over all histories  and 
$p_{\alpha}$ is the
probability for the occurence of the history $\alpha$.  Such histories do not 
necessarily satisfy the GMHG consistency requirement.
Thus, defining the trace as the diagonal sum of expectations over all
histories (which are
conventionally denoted inside the tensor product as $\phi^{\alpha}$), 
in the sense of Isham's complteteness,
  $$tr(\rho)=\sum_{\grave
\alpha}(\phi^{\grave \alpha},\rho \phi^{\grave \alpha})$$ If we take a subset of
histories $\grave \alpha$ to correspond to the $\alpha$'s that occur in $\rho$,
and the remainder orthogonal to these, we see that \begin{equation} \label{e14} 
tr(\rho)=\sum_{\alpha}p_{\alpha}\prod_i|(\phi^{\alpha_i},\phi^{\alpha_i})|^2=
\sum_{\alpha}p_{\alpha}=1 \end{equation} 
Such a density operator is consistent with the notion of the
Lax-Phillips state, since in this theory each pure state corresponds to a
quantum mechanical history.  A decoherence functional for such state
defined by
        $d(\alpha, \grave \alpha') = tr (C_{\alpha} \rho C^{\dagger}_{\grave \alpha})$
satisfies the consistency conditions for any pair of histories satisfying the
condition \ref{e10}, i.e., 
 \begin{equation} \label{e15} tr(C_{\alpha}\rho C^{\dagger}_{\grave
\alpha})=\sum_{\beta}\sum_{\alpha^"}p_{\alpha^"}\prod_i
(\phi^{\beta_i},P_{\alpha_i}P_{\alpha^"_i}P_{\grave \alpha_i}
\phi^{\beta_i})
\end{equation}
Choosing a subset of histories $\beta$ to coincide with $\alpha$,  and the
remainder orthogonal to these,  we see that $$ tr(C_{\alpha}\rho C^{\dagger}_{\grave
\alpha})=\sum_{\alpha^"}p_{\alpha^"}\prod_i\delta_{\alpha_i
\alpha^"}\delta_{\alpha^"\grave \alpha_i}=p_{\alpha}\delta_{\alpha
\grave \alpha}$$ i.e., the analog of the GMHG condition for consistent 
histories is formally 
satisfied with these
definitions.  Furthermore,  for an observable $A$ {\it defined on the
 space of histories} (as for operators on the Lax-Phillips space $\hat {H}$)
 \begin{equation} \label{e16} tr(A\rho)=\sum_{\alpha}\sum_{\grave \alpha}p_{\grave
\alpha}\prod_i(\phi^{\alpha_i},AP_{\grave \alpha_i}\phi^{\alpha_i}), 
\end{equation} and again taking a subset of histories $\alpha$ to coincide with
the set occuring in $\rho$, and the remainder orthogonal to these, one finds
\begin{equation} \label{e17} tr(A\rho)=\sum_{\alpha}p_{
\alpha}\prod_i(\phi^{\alpha_i},A\phi^{\alpha_i}) \end{equation}  
 We remark that the Lax-Phillips Hilbert space contains elements that
represent resonances, and which evolve according to exact semigroup laws
(and hence correspond to irreversible processes).  This property cannot be
achieved in the framework of the usual quantum theory using the
Wigner-Weisskopf formulation \cite{Wigner} of the description of an unstable 
system. \par
The projection operators
$C_{\alpha}$ are proper projection operators in $\hat{H}$. With suitable
conditions, identifying incoming and outgoing subspaces in $\hat H$
which are stable  under the action of $U(\tau)$ for $\tau$ positive and 
negative, 
respectively, the quantum Lax-Phillips theory \cite{Lax} provides a rigorous
framework for describing irreversible processes  (semigroup evolution of a
subspace of $\hat H$)   here seen as an evolution of  the GMHG histories \par    
The 
action of $U(\tau)$ 
is generated by  a self adjoint generator \begin{equation} K=s-\lim_{\tau \to 0}{1\over 
\tau}(U(\tau)-I) \label{e18}\end{equation}  For example, the Flesia-Piron model 
 \cite{Horwitz} is 
\begin{equation}K=H-i\partial_t,  \label{e19} \end{equation} where $H$ is a
(possibly $t$-dependent) Hamiltonian operator defined on the auxiliary spaces. 
 This generalized 
generator $K(q,p,t,E)$ (which may depend on the variable 
$t$ but not on the laboratory parameter time $\tau$; $E$ is represented by
$i\partial_t$) satisfies the 
following generalized equation \cite{Horwitz} \begin{equation} \label{e20}
i{\partial \over \partial \tau}(U(\tau)\phi)_t =(U(\tau)K\phi)_t=V_{t-\tau}
(\tau)(K\phi)_{t-\tau} 
=V_{t-\tau}(\tau)(H_{t-\tau}\phi_{t-\tau}-i\partial_t\phi_{t-\tau}) 
\end{equation}
Using Equations (\ref{e3})-(\ref{e6}) we write the left hand side of 
Eq (\ref{e20}) as 
\begin{equation} \label{e21}
  i{\partial \over \partial \tau}(U(\tau)C_{\alpha}\phi) 
   =  i{\partial \over \partial \tau}(V_{t_1}(\tau)P_{\alpha_1}(t_1)h_{t_1},
 V_{t_2}(\tau)P_{\alpha_2}(t_2)h_{t_2}\ldots) 
\end{equation}
In the case that the state $C_{\alpha}\phi$ is stationary under the evolution
$U(\tau)$, so that 
\begin{equation} \label{e22} V_{t_i}(\tau)P_{\alpha_i}(t_i)h_{t_i}=
P_{\alpha_i}(t_i)h_{t_i},  \end{equation}  one obtains,  in the special case 
of Eq (\ref{e19}),  the Schr\"oedinger 
equation \begin{equation} \label{e23} 
H_{t_i}\phi_{t_i}=i\partial_{t_i}\phi_{t_i}   \end{equation} at every $t_i$.
 We now show that for a special choice of evolution relating
successive sequences \begin{equation} \label{e24} 
h_{t_k+\delta \tau}=V_{t_k}(\delta \tau)h_{t_k}= h_{t_k}(\delta \tau), 
\end{equation}  
with $\delta \tau=\frac{T}{n}$, $T$ is the total span $t_1, t_2, \ldots t_n$, 
we obtain the stability of the corresponding  $C_{\alpha}\phi$ by a
mechanism analogous to the dynamical Zeno effect \cite{Aharonov,Facchi}. 
Since $V_{t_k}(\delta \tau)$ is generated on the auxiliary space ${\cal H}_{t_k}$ 
by a self adjoint operator $H_{t_k}$, for $n$ large Eq (\ref{e24}) may be
written to second order in $\delta \tau$ as \begin{equation} \label{e25} 
(1-i\delta \tau H_{t_k}-\frac{\delta
\tau ^2}{2}H_{t_k}^2)h_{t_k}=\grave h_{t_k+\delta \tau}, \end{equation}  so that for $\delta \tau \to 0$ the
Schr\"oedinger equation (\ref{e23}) is satisfied (here we take $\delta \tau=\delta
t$). Applying the definition (\ref{e12}) of the tensor  product 
we see that,  identifying  $\grave h_{t_k+\delta \tau} \equiv h_{t_k}(\delta \tau)$ as
associated with the component $t_k$,  \begin{equation} \label{e26} <\!U(\delta \tau)\phi, \phi\!>=
\prod_k<\!(1-i\delta \tau H_{t_k}-\frac{\delta
\tau ^2}{2}H_{t_k}^2)h_{t_k},  h_{t_k}\!> \end{equation}  
It then follows that \begin{equation} \label{e27} 
|<\!U(\delta \tau)\phi, \phi\!>|^2=\prod_k||h_{t_k}||^4(1-\delta \tau^2 \Delta
H_{t_k}^2), \end{equation} where, \begin{equation} \label{e28} 
\Delta H_{t_k}^2=
\frac{<\!h_{t_k}, H^2_{t_k}h_{t_k}\!>_{{\cal H}_{t_k}}}{||h_{t_k}||^2}-
\frac{<\!h_{t_k}, H_{t_k}h_{t_k}\!>^2_{{\cal H}_{t_k}}}{||h_{t_k}||^4}
\end{equation} The product to second order is  \begin{equation} \label{e29} 
|<\!U(\delta \tau)\phi,\phi\!>|^2=(\prod_k||h_{t_k}||^4)(1+\frac{T^2}{n^2}\sum_k
 \Delta H_{t_k}^2) \end{equation} The second term vanishes in the limit of $n \to
 \infty$ if the dispersions $\Delta H_{t_k}$ are finite at each $k$. 
 The GMHG scalar product for this evolution constructs the sequence
corresponding to the well-known Zeno phenomenon in the stabilization of states
 by successive measurement. Moreover,  this
result demonstrates that the set of consistent histories
constructed by successive measurement is associated with the Schr\"odinger
relation between successive states in the Lax-Phillips foliation.
  \par 
 In this calculation we have concluded that the GMHG evolution is stationary if
 the sequence ${h_{t_k}}$ is determined by Schr\"oedinger evolution in the
 Flesia-Piron model \cite{Horwitz}. However, if we choose
 an {\it arbitrary} sequence ${h_{t_k}}$ and $U(\delta \tau)$ induces Schr\"oedinger
 evolution at each step $\delta \tau$, the relation (\ref{e26}) remains valid,
 independently of whether the sequence ${h_{t_k}}$ itself corresponds to a
 Schr\"oedinger sequence. The conclusion (\ref{e29}) remains, and we see the
 resulting GMHG stability corresponds to the dynamic Aharonov-Vardi
 \cite{Aharonov} Zeno effect. The static Zeno effect follows if the sequence
 ${h_{t_k}}$ is chosen to be a set of identical states, i.e., $h_{t_k}=h_{t_0}$.
 \bigskip    \bibliographystyle{plain}

\end{document}